\documentclass[preprint,
    aps,prd,
    amsmath,amssymb,
    superscriptaddress,
    twoside,
    nofootinbib,
    showkeys,
    floatfix,
    longbibliography
]{revtex4-2}

\usepackage{graphicx}
\usepackage{dcolumn}
\usepackage{bm}
\usepackage{physics}
\usepackage{lineno}
\usepackage{xcolor}
\usepackage{cancel}
\usepackage{float}
\usepackage[pdfstartview=XYZ,
    bookmarks=true,
    colorlinks=true,
    linkcolor=blue,
    urlcolor=blue,
    citecolor=blue,
    bookmarks=true,
    linktocpage=true, % makes the page number as hyperlink in table of content
    hyperindex=true
]{hyperref}
\usepackage{tikz}

\newcommand{\Lag}{\mathcal{L}}
\newcommand{\Hc}{\text{H.c.}}
\DeclareMathOperator{\diag}{diag}

\begin{document}

%\preprint{}

%%ORCID LOGO%%%
\definecolor{lime}{HTML}{A6CE39}
\DeclareRobustCommand{\orcidicon}{%
	\begin{tikzpicture}
	\draw[lime, fill=lime] (0,0) 
	circle [radius=0.16] 
	node[white] {{\fontfamily{qag}\selectfont \tiny ID}};
	\draw[white, fill=white] (-0.0625,0.095) 
	circle [radius=0.007];
	\end{tikzpicture}
	\hspace{-2mm}
}

\foreach \x in {A, ..., Z}{\expandafter\xdef\csname orcid\x\endcsname{\noexpand\href{https://orcid.org/\csname orcidauthor\x\endcsname}
	{\noexpand\orcidicon}}
}

\newcommand{\orcidauthorA}{0000-0002-5757-8810}
\newcommand{\orcidauthorB}{0000-0002-4574-9336}
\newcommand{\orcidauthorC}{0000-0003-3037-3708}
\newcommand{\orcidauthorD}{0000-0003-2104-8460}
\newcommand{\orcidauthorE}{0000-0001-5279-8438}

\title{Hidden physics in pion and some other meson decays}
\author{M. M. Guzzo\orcidA{}}
    \email{guzzo@ifi.unicamp.br}
\author{L. J. F. Leite\orcidB{}}
    \email{lfleite@ifi.unicamp.br}
\author{S. W. P. Novelo\orcidC{}}
    \email{wnovelo@ifi.unicamp.br}
\author{O. L. G. Peres\orcidD{}}
    \email{orlando@ifi.unicamp.br}
\affiliation{
Instituto de F\'isica Gleb Wataghin - UNICAMP, 13083-859, Campinas-SP, Brazil
}

\author{V. Pleitez\orcidE{}}
    \email{v.pleitez@unesp.br}
\affiliation{Instituto  de F\'isica Te\'orica, Universidade Estadual Paulista, \\R. Dr. Bento Teobaldo Ferraz 271, Barra Funda, S\~ao Paulo - SP, 01140-070\\ Brazil.
}

\date{
03/09/2023
%\today
}

\begin{abstract}
It has been commonly assumed that pseudoscalar contributions to the leptonic decay of charged mesons, like pions and kaons, is strongly constrained due to the helicity suppression present in the ratio $R_{l/l'} = \Gamma(P \rightarrow l\nu[\gamma])/\Gamma(P \rightarrow l'\nu[\gamma])$, where $P$ are the charged pseudoscalar meson and $l,l' = e,\mu,\tau$. Here we show that if the effective couplings are proportional to the corresponding charged lepton masses (and also the PMNS matrix), the constraints from $R_{l/l'}$  are entirely avoided, and a rather new large allowed region is permitted in the parameter space. In the case of the electron, we found a non-trivial region in the range $10^{-4}\lesssim (G^\eta/G_F) \lesssim 10^{-3}$, where $G^\eta$ is the effective pseudoscalar coupling associated with a novel charged scalar field, $\eta$, and $G_F$ is the Fermi constant. Furthermore, we show that this dependence of the pseudoscalar couplings on the charged lepton masses can naturally be associated with a critical class beyond the standard model physics, namely models without (leptonic) flavor-changing neutral currents in the scalar sector. The most known examples are the models that satisfy the so-called Glashow-Weinberg-Paschos theorem. Finally, we also point out that, in those cases, the decay rate is degenerated with the Standard Model prediction, possibly hiding the new physics effects in those decays.

\end{abstract}

\maketitle

%%%%% INTRODUCTION %%%%%
\section{Introduction}
\label{sec:Intro}

In most possible new physics Beyond the Standard Model (BSM), even when considering their minimal versions, a more complex scalar sector is encountered than the simple neutral Higgs present in the particle spectrum of the Standard Model (SM)~\cite{Rosner:2002xi}. Even in the context of the ${\rm SU_L}(2) \times {\rm U_Y}(1)$ gauge symmetry, nothing limits the number of scalar fields. However, at least a single doublet is necessary for the usual spontaneous symmetry-breaking pattern. Thus, one cannot rule out the possibility that extra scalars, heavier than the observed Higgs (or lighter, but with sufficiently weak couplings), exist. Moreover, many mechanisms to generate neutrino masses require additional scalars ~\cite{Cheng:1980qt,Schechter:1980gr,Branco:2011iw}. Nevertheless, in principle, such particles can solve some anomalies in high-energy experiments, like those in B-meson decays \cite{Celis:2012dk,Schacht:2020qot} or the muon anomalous magnetic moment~\cite{Czarnecki:2001pv, Muong-2:2021ojo}, for example.

If those extra scalars exist, they may modify several well-known processes, such as leptonic pion decay. Such a process has an astonishing agreement between the experimental results and the SM theoretical calculations, often used as a hallmark of the weak interactions' V-A structure. Moreover, its helicity suppression explains the dominant decay in muons (99.99\%), not electrons. Therefore, strong constraints on new physics (especially from pseudoscalar interactions) are possible in this decay~\cite{Herczeg:1994ur,Herczeg:1995kd,2001PrPNP46413H,Campbell:2003ir,Campbell:2008um,Dobrescu:2008er,Cirigliano:2011ny,Bhattacharya:2011qm,Celis:2012dk,Cirigliano:2013xha,Crivellin:2013wna,Enomoto:2015wbn,
Gonzalez-Alonso:2016etj,
Barranco:2016njc,
Wei:2017ago,Banelli:2018fnx,Botella:2018gzy,
Cirigliano:2018dyk,Zhang:2019tcs,BESIII:2019vhn,Li:2020wxi,Han:2020pff,Cornet-Gomez:2021lep,Diaz-Cruz:2020pjf,
Leng:2020fei,Fleischer:2021yjo,Falkowski:2021bkq,Bryman:2021teu,Davoudiasl:2021nfv,Ansarifard:2021dju,Aloni:2022ebm,Breso-Pla:2023tnz}. 

The main goal of this work is to stress that this parameter space region in the leptonic decays with pseudoscalar interactions could be hidden in well-motivated scenarios. Theoretically, in models with a Glashow-Weinberg-Paschos (GWP) mechanism implemented and phenomenological using the helicity-suppressed ratio as observable, we automatically cancel new physics effects, rendering these well-known tests for charged scalar new physics ineffective. Therefore, we choose to use a fundamental Lagrangian, in order to analyze this kind of decays under the model building approach.

\section{Charged Meson Decay}
\label{sec:MesonDecay}

Consider the leptonic decay of charged mesons, $P^+ \rightarrow \ell^+ \nu [\gamma]$ (henceforth denoted by $P_{\ell 2}$), in the presence of a novel pseudoscalar interaction among the SM fermions. The low-energy effective Lagrangian, in this case, is given by
\begin{equation}
\label{eqn:EffectiveLagrangian}
    -\Lag_\text{eff} = \frac{G_F}{\sqrt{2}} V_{ij} \left\{\bar{u}_i \gamma_\mu (1-\gamma_5)d_j \cdot \bar{\ell}_l \gamma^\mu (1-\gamma_5)\nu_l + \frac{\mathcal{G}^\eta_{ij,ll'}}{G_F} (\bar{u}_i \gamma_5 d_j) \cdot \bar{\ell}_l(1-\gamma_5)\nu_{l'} \right\}+ \Hc,
\end{equation}
where $G_F$ is the tree-level Fermi constant, $V_{ij}$ is an element of the Cabbibo-Kobayashi-Maskawa (CKM) matrix~\cite{Cabibbo:1963yz,Kobayashi:1973fv}, and $\mathcal{G}^\eta$ is the effective coupling matrix of the new four-fermion interaction in the neutrino interaction basis. Notice that we also factored out a CKM matrix element from the new physics term. Each meson $P$ fixes the corresponding quark indices $i,j$, while the lepton indices assume the values $l,l' = e,\mu,\tau$. The SM effective contribution, mediated by a $W$-boson exchange, corresponds to the first term between curly brackets in the above equation. Unless explicitly stated otherwise, all repeated indices are summed over throughout this paper. 

As is well known, the left-handed neutrino fields $\nu_{lL} = \tfrac{1}{2}(1-\gamma_5)\nu_l$ are actually a linear combination of the mass eigenstates $\nu_{kL}$,
\begin{equation}
    \nu_{lL} =  U_{lk} \nu_{kL},
\end{equation}
where $U$ is the Pontecorvo-Maki-Nakagawa-Sakata (PMNS) mixing matrix~\cite{Pontecorvo:1957cp,Maki:1962mu}. Therefore, in the neutrino mass basis, the effective matrix coupling for the new interaction is given by
\begin{equation}
\label{eqn:G-Mass-to-Flavor}
    G^\eta_{ij,lk} \equiv \mathcal{G}^\eta_{ij,ll'} U_{l'k}.
\end{equation}

The most simple realization of the new effective operator in Eq.~\eqref{eqn:EffectiveLagrangian}, and the one we will be interested in, is through the Yukawa interaction of the SM fields with a new charged scalar field $\eta$
\begin{equation}
\label{eqn:GeneralLagrangian}
    -\Lag = \bar{u}_{i}\left(c^s_{ij} + c^p_{ij}\gamma_5 \right) d_{j} \eta^+ + X_{lk} \bar{\ell}_{lR} \nu_{kL} \eta^- + \Hc,
\end{equation}
where $c^s$ and $c^p$ are, respectively, the scalar and pseudoscalar Yukawa couplings in the quark sector and $X$ is the matrix of Yukawa couplings in the lepton sector. Matching the interactions in Eqs.~\eqref{eqn:EffectiveLagrangian} and~\eqref{eqn:GeneralLagrangian}, and using the relation in Eq.~\eqref{eqn:G-Mass-to-Flavor}, we have that
\begin{equation}
\label{eqn:EffectiveScalarOperator}
    \frac{G^\eta_{ij,lk}}{\sqrt{2}} =  \frac{c^p_{ij}}{V_{ij}} \frac{X_{lk}}{m_\eta^2},
\end{equation}
with $m_\eta$ being the mass of the new scalar field.

To calculate the amplitude for $P_{\ell 2}$, the following matrix elements are needed (all other matrix elements are null for pseudoscalar mesons)
\begin{align}
\label{eqn:Axial-Vector-Current}
    \mel{0}{\bar{u}_i\gamma^\mu \gamma_5 d_j}{P^+(k)} = i k^\mu f_P, \quad \mel{0}{\bar{u}_i \gamma_5 d_j}{P^+(k)} = i \tilde{f}_P,
\end{align}
where $k^\mu$ is the meson linear momentum, $f_P$ is the corresponding meson decay constant, and $\tilde{f}_P$ and $f_P$ are related by the identity~\cite{Bernard:2006gx}
\begin{equation}
\label{eqn:FactorBP}
    \frac{\tilde{f}_P}{f_P} = \frac{m_P^2}{m_{u_i} + m_{d_j}} \equiv B_P,
\end{equation}
with $m_P$ being the charged meson mass, and $m_{u_i}$ and $m_{d_j}$ the bare masses of its constituent quarks. Notice that, due to the quark masses, the $B_P$ factor depends on both the renormalization scale and scheme, that is, $B_P = B_P(\mu)$~\cite{Cvetic:1997zd,Cvetic:1998uw,Grimus:2004yh,Campbell:2008um,Buras:2010mh,Braeuninger:2010td,Ahn:2010zza,Cvetic:1997zd,Bhattacharya:2011qm}. Our analysis were entirely made using $\mu = 2\,\text{GeV}$ in the $\overline{\text{MS}}$ scheme, and we only consider the pseudoscalar contribution (with magnitude $\varepsilon_P$). The only exception being the analysis of the $B$-meson, where we first calculated using $\mu = m_b$, and then we go to $\mu = 2\,\text{GeV}$, using the renormalization group equations. From Ref.~\cite{Gonzalez-Alonso:2017iyc}, we know that the pseudoscalar coupling changes a little when we change scales from $\mu = 2\,\text{GeV}$ to $\mu = m_b$, namely $\varepsilon_P(\mu = 2\,\text{GeV}) \approx 1.178\,\varepsilon_P(\mu = m_b)$. Another effect we might expect by the running of renormalization group equation is the generation of new couplings in a different scale. In our case, it can be showed that the scalar and tensorial contributions will be suppressed, $\varepsilon_S (\mu = 2\,\text{GeV})/\varepsilon_P (\mu = m_b) \approx 10^{-8}$ and $\varepsilon_T (\mu = 2\,\text{GeV})/\varepsilon_P (\mu = m_b) \approx 10^{-5}$, which justify our use of only pseudoscalar interactions.

With the above assumptions, the total decay rate for $P_{\ell 2}$, in the meson rest frame, is given by
\begin{equation}
\label{eqn:Total-Decay-Rate}
    \Gamma_l = \Gamma_l^{\rm SM}\times\left(1 + \Delta_l\right),
\end{equation}
where
\begin{equation}
    \Gamma_l^{\rm SM} = r_l \frac{G_F^2}{8\pi m_P^3} f_P^2 V_{ij} m_l^2 (m_P^2 - m_l^2)^2
\end{equation}
corresponds to the usual Standard Model rate, including radiative corrections $r_l$ for soft photons~\cite{Workman:2022ynf,Cirigliano:2007ga},
\begin{equation}
\label{eqn:Delta-L}
    \Delta_l = \frac{B_P}{m_l}\sum_{k=1}^{3} \Bigg[ \left(\frac{\abs{G^\eta_{l k}}}{G_F}\right)^2 \frac{B_P}{m_l} - 2\Re\left(U_{lk} \frac{(G^\eta)_{lk}^*}{G_F}\right) \Bigg], \quad (l=e,\mu,\tau)
\end{equation}
quantifies the presence of new physics beyond the SM, and $m_l$ is the mass of the final state-charged lepton. For simplicity, we omitted the quark indices in $G^\eta$, since they are fixed for each meson. Terms proportional to the neutrino masses were neglected, and for each charged lepton state, we summed over all the active neutrino mass eigenstates ($k=1,2,3$). In Eq.~\eqref{eqn:Delta-L}, the first term inside square brackets comes purely from the pseudoscalar interaction, and the latter corresponds to the interference between the SM contribution and the new interaction.

Experimental results require $\abs{\Delta_l} \ll 1$. For example, for pion decay, we must have $\abs{\Delta_l} \lesssim 10^{-3}$, using the central value of the experimental result and the current experimental uncertainties. Therefore, an agreement of the experimental data and the SM prediction can be possible in BSM scenarios if we set $\Delta_l\equiv 0$, rendering the new physics contributions hidden for these observable. For example, most analyses present in the literature, as in~\cite{Bhattacharya:2011qm,Herczeg:1994ur,Herczeg:1995kd}, assume that, for each meson $P$, the effective couplings $G^\eta_{lk}$ has a similar size for all charged leptons and neutrino states. Therefore, due to the enhancement factor $B_P/m_l$ present in Eq.~\eqref{eqn:Delta-L}, the most sensible channel for new physics is the decay with electrons in the final state. Here, in this paper, we make two different assumptions, namely (i) that the effective coupling $G^\eta_{lk}$ depends directly on the charged lepton masses, and (ii) that the neutrino flavor is conserved. That is, \textit{we assume} that, in the neutrino mass basis,
\begin{equation}
\label{eqn:AnsatzG}
    G^\eta_{lk} = a_P m_l U_{lk}, \qq{(no sum in $l$)}
\end{equation}
where $a_P$ depends only on the decaying charged meson and the scalar field $\eta$ that mediates the interaction. Although we, again, omitted the quark indices in the above expression, a subscript $P$ was added to $a_P$ to remind the reader that this quantity can be different to each meson. Using Eq.~\eqref{eqn:G-Mass-to-Flavor}, the corresponding expression in the neutrino flavor basis is given by
\begin{equation}
\label{eqn:AnsatzGFlavor}
    \mathcal{G}^\eta_{ll'} = a_P m_l \delta_{ll'}, \qq{(no sum in $l$)}
\end{equation}
where $\delta_{ll'}$ is the usual Kronecker delta.
 
Using Eq.~\eqref{eqn:AnsatzG}, and the unitarity of the PMNS matrix, the new physics contribution given in Eq. \eqref{eqn:Delta-L} becomes
\begin{equation}
\label{eqn:Delta_aP}
    \Delta_l = B_P \Bigg[ \left(\frac{\abs{a_P}}{G_F}\right)^2 B_P - 2\Re\left(\frac{a_P^*}{G_F}\right) \Bigg] \equiv \Delta,
\end{equation}
Furthermore, all the charged leptons are equal, although it can differ for each meson. 

In the literature, in order to avoid uncertainties coming from $f_P$, the ratio
\begin{equation}
\label{eqn:Br-Ratio}
    R_{l/l'} = \frac{\Gamma(P \rightarrow l \nu [\gamma])}{\Gamma(P \rightarrow l' \nu [\gamma])} = R^\text{SM}_{l/l'}\left(\frac{1+\Delta_l}{1 + \Delta_{l'}}\right), \quad (l,l' = e,\mu,\tau),
\end{equation}
is commonly used to constrain new physics~\cite{Bhattacharya:2011qm,Herczeg:1994ur,Herczeg:1995kd}. However, as we just saw, for our solution in Eq.~\eqref{eqn:AnsatzG}, $\Delta$ is independent of the final lepton states. Therefore, all new contributions for these ratios are automatically canceled, that is 
\begin{equation}
    R_{l/l'}  = R_{l/l'}^\text{SM},
\end{equation}
irrespective of the value of $a_P$. With this, the usual helicity suppression in meson decays within the SM is recovered, and the strong constraints usually assumed to come from these observables will not apply. Although the fact that, with an \textit{Ansatz} as in Eq.~\eqref{eqn:AnsatzG}, the ratios coincide with its SM value had already been mentioned in the literature cited above, no statistical analysis of the allowed parameter space using the individual rates has ever been performed, to our knowledge, being also a novelty of this work.

Besides automatic canceling new contributions for the ratio $R_{l/l'}$, the effective coupling of Eq.~\eqref{eqn:AnsatzG} can be chosen such that the effects on the individual rates $\Gamma_l$ also vanish, rendering the new physics contribution on the leptonic decays of pseudoscalar mesons completely hidden. Then, taking  $\Delta = 0$, we find that $a_P$ must satisfy
\begin{equation}
\label{eqn:MasterEquation}
    \abs{a_P}^2 - \frac{2 G_F}{B_P} \Re(a_P) = 0.
\end{equation}

A trivial solution for Eq.~\eqref{eqn:MasterEquation} would be $a_P = 0$, and perturbations around this solution correspond to a weakly coupled scalar, either because the associated Yukawa couplings in Eq.~\eqref{eqn:EffectiveScalarOperator} are small or because the scalar has a huge mass. However, other non-trivial solutions are possible and will be discussed in the following subsections. However, other non-trivial solutions are possible and will be discussed in the following subsections. Moreover, as we will show in Section~\ref{sec:GWP}, such a non-trivial solution can naturally arise in models where flavor-changing neutral currents (FCNC) are absent in the scalar sector at tree level. In such models, constraints coming from charged meson decays can be less stringent than one would first assume. Nevertheless, before we enter such a discussion, let us analyze the allowed parameter space for a non-trivial solution of Eq.~\eqref{eqn:AnsatzG}.

\subsection{Real solutions}
\label{sec:Real-AP}

If we assume that $a_P$ is a (non-zero) real parameter, Eq.~\eqref{eqn:MasterEquation} simplifies to
\begin{equation}
\label{eqn:Real-AP-Solution}
    \frac{a_P}{G_F} = \frac{2}{B_P}.
\end{equation}

\begin{figure}[htp]
    \centering
    \includegraphics[width=0.5\textwidth]{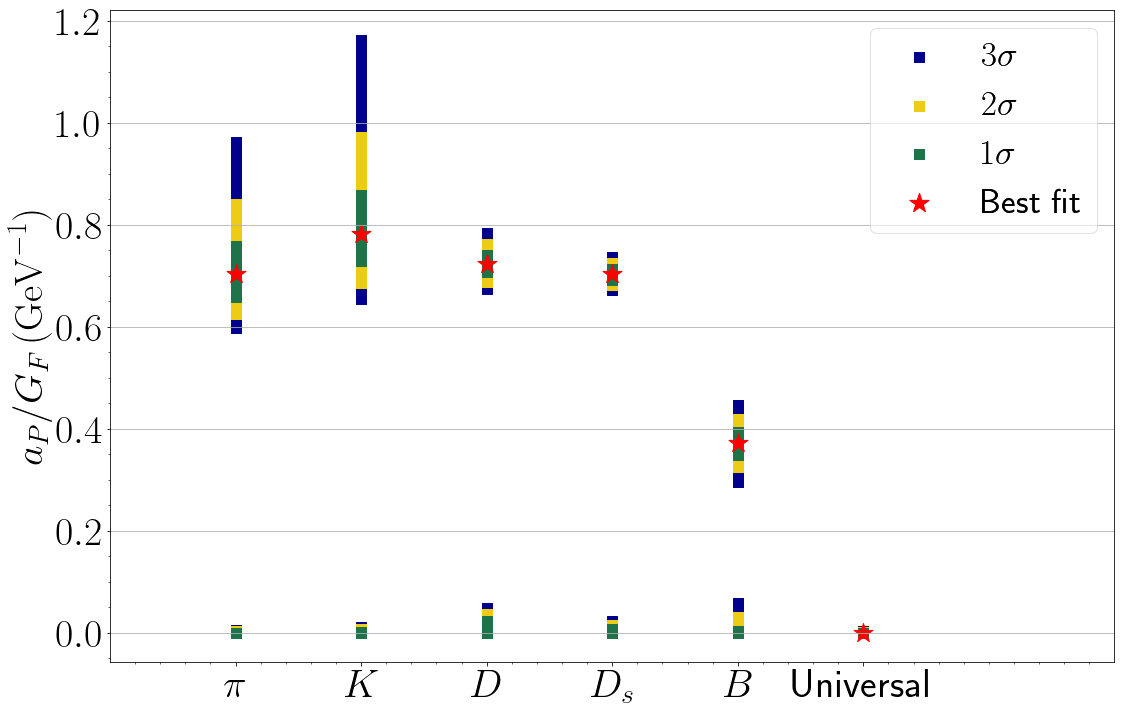}
    \caption{Allowed region for new physics for each pseudoscalar meson $\pi$, $K$, $D$, $D_s$ and $B$, up to $3\sigma$, assuming a real parameter $a_P$ in Eq.~\eqref{eqn:AnsatzG}. We also show the analysis fitting the experimental data (as detailed in Appendix \ref{sec:StatisticalAnalysis}) for all five mesons considering a universal parameter on the quark sector. The allowed region, in this case, is hidden behind the best-fit marker.}
    \label{fig:Real-AP-Mesons}
\end{figure}

Figure~\ref{fig:Real-AP-Mesons} shows the allowed region for $a_P$ constrained by the experimental values of the decay rates for each meson $P = \pi, K, D, D_s$ and $B$. As can be seen, two distinct regions arise. The first one, with $a_P \approx 0$, is compatible with a full dominant SM process, as discussed above. The second region, with $a_P/G_F$ values of the order of $\text{GeV}^{-1}$, corresponds to the non-trivial solution given in Eq.~\eqref{eqn:Real-AP-Solution}, and will be called, henceforth, the \textit{non-trivial region}. A non-trivial region not only appears as a possible solution to all the mesons considered but is slightly preferred by our statistical analysis, as indicated by the best points. 

As seen from Figure~\ref{fig:Real-AP-Mesons}, the non-trivial regions occur within the same range for all the mesons except for the B meson. Since the quark condensates are approximately equal for the lightest quarks, the factor $B_P$, defined in Eq.~\eqref{eqn:FactorBP}, turns out to be almost identical for the lightest mesons. The deviations in $B_P$ are larger for mesons containing heavier quarks like the $B$-meson \cite{Gonzalez-Alonso:2016etj}. This feature explains why the non-trivial region for the $B$-meson is considerably lower than the other mesons. In the last column of Figure \ref{fig:Real-AP-Mesons}, we present a global fit for the leptonic decays of the five considered mesons considering a single universal parameter independent of the meson type, that is, 
\begin{equation}
    a_\pi = a_K = a_D = a_{D_s} = a_B.    
\end{equation}
No  non-trivial  universal solution was found in this last analysis considering all five mesons, and only solutions around the SM value appear. However, if we consider an ``universal'' solution for $\pi$ and $K$, that is, $a_\pi=a_K$, the ratio $\pi_{\mu 2}/K_{\mu 2}$, used in some non-standard searches analyses in the literature to reduce lattice uncertainties~\cite{Cirigliano:2011tm,Jung:2010ik}, cancels the new physics contribution for any value of $a_P=a_\pi=a_K$, similarly from what happens in the case of $R_{l/l'}$. Therefore, we see that it can avoid strong constraints from this observable in this case. It is also possible to note from Figure~\ref{fig:Real-AP-Mesons} that the parameter space for pion and kaon overlap at $1\sigma$ even at the non-trivial region, so an ``universal'' solution of this kind is not excluded by these decays alone.

As a numerical example, for the particular case of pion decay, and using the best-fit point present in the non-trivial region in Figure \ref{fig:Real-AP-Mesons}, we have that the effective coupling $G^\eta_{lk}$ is given by
\begin{equation}
\label{eqn:G-Pion-Mass}
    \frac{G^\eta_{lk}}{G_F} \approx 
    \begin{pmatrix}
        3 \times 10^{-4} & 2  \times 10^{-4} & 5 \cdot 10^{-5} \\
        - 3  \times 10^{-2}           & 4  \times 10^{-2}             & 5  \times 10^{-2}  \\
      4  \times 10^{-1}          & 8  \times 10^{-1}            & 8  \times 10^{-1} \\
    \end{pmatrix}
\end{equation}
in the neutrino mass basis, or
\begin{equation}
\label{eqn:G-Pion-Flavor}
    \frac{\mathcal{G}^\eta_{l\,l'}}{G_F} \approx 
    \begin{pmatrix}
        4 \times 10^{-4} & 0    & 0 \\
        0               & 7  \times 10^{-2} & 0 \\
        0               & 0    & 1.2 \\
    \end{pmatrix}
\end{equation}
in the respective neutrino flavor basis. We point out that, in the neutrino mass basis, for any fixed line, each column element differ from the others due to the presence of the PMNS matrix in our assumption given in Eq.~\eqref{eqn:AnsatzG}, which includes the negative signs appearing in Eq.~\eqref{eqn:G-Pion-Mass}. Finally, in both neutrino bases, the hierarchy between every two lines directly results from the charged lepton masses dependence.

\subsection{Complex solutions}
\label{sec:Complex-AP}
\begin{figure}[htp]
    \centering
    \includegraphics[width=.5\textwidth]{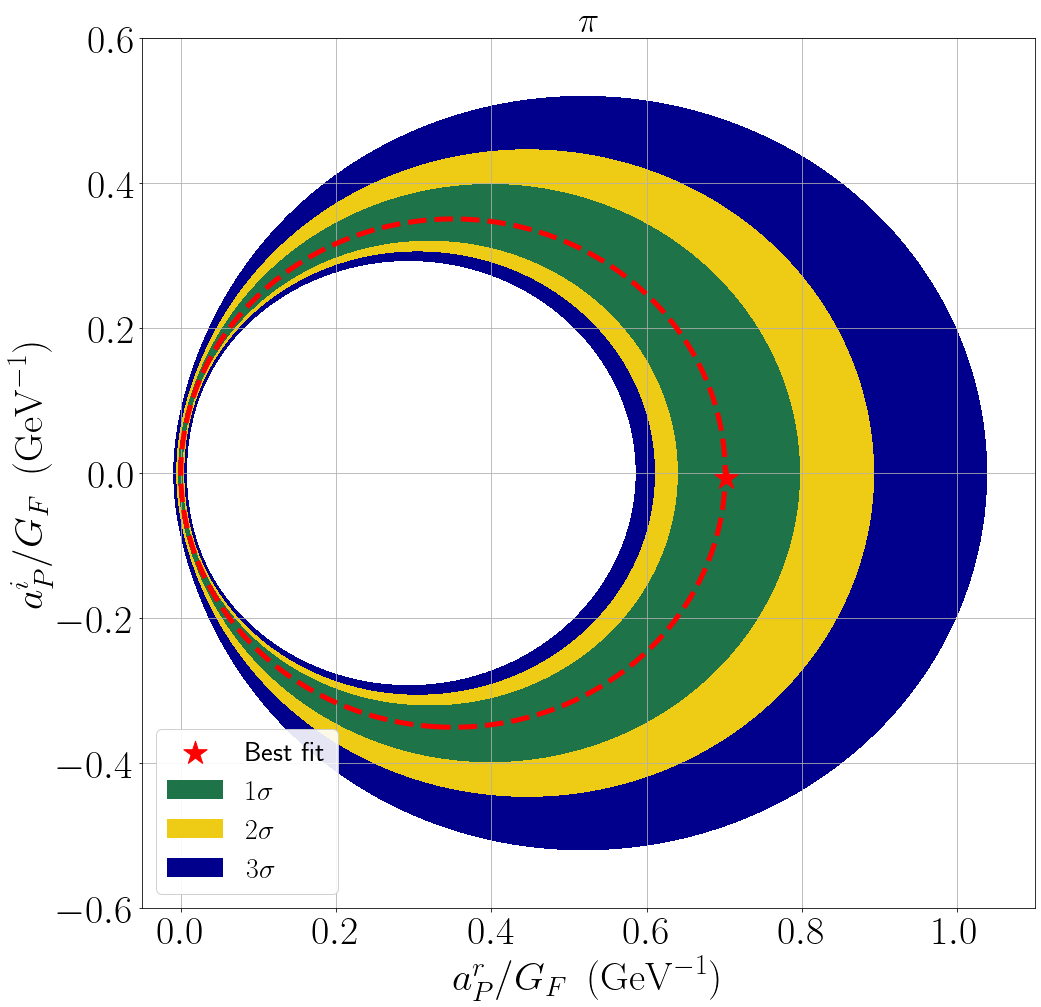}
    \caption{Allowed region for the meson $\pi$, up to $3\sigma$, assuming a complex parameter $a_P$. The red dashed curve corresponds to the fine-tuned (exact) solution of Eq.~\eqref{eqn:Complex-AP-Solutions}. However, as can be seen from the above figure, a larger allowed region appears beyond this fine-tuned solution.}
    \label{fig:Complex-AP-Pion}
\end{figure}

In general, the $a_P$ parameter will be complex. In this case, Eq.~\eqref{eqn:MasterEquation} can be written as
\begin{equation}
\label{eqn:Complex-AP-Solutions}
    \left(\frac{a_P^r}{G_F} - \frac{1}{B_P}\right)^2 + \left(\frac{a_P^i}{G_F}\right)^2 = \left(\frac{1}{B_P}\right)^2
\end{equation}
where $a_P^r = \Re(a_P)$ and $a_P^i = \Im(a_P)$ are the real and imaginary parts of $a_P$, respectively. Eq.~\eqref{eqn:Complex-AP-Solutions} describes a circle of radius $B_P^{-1}$, centered in $(B_P^{-1}, 0)$ in the $a_P^r/G_F \times a_P^i/G_F$ plane. From Eq.~\eqref{eqn:Complex-AP-Solutions}, it is easy to see that, for $a_P^i = 0$, we can recover the real non-trivial solution of Eq.~\eqref{eqn:Real-AP-Solution}, while for $a_P^r = 0$ the only possible solution is $a_P^i = 0$.

Figure~\ref{fig:Complex-AP-Pion} shows the allowed region for a complex $a_P$ for the case $\pi_{l2}$. As can be seen, apart from the uncertainties, a connected circular region is encountered. Now, solutions compatible with a full-dominant SM contribution correspond to small perturbations around the point $(0,0)$ in the $a_P^r/G_F \times a_P^i/G_F$ plane, while the rest of the circle corresponds to a non-trivial solution of Eq.~\eqref{eqn:Complex-AP-Solutions}.

\section{Models with flavor conserving neutral currents}
\label{sec:GWP}

In the previous section, we propose and study a non-trivial solution for the new physics, given in Eq.~\eqref{eqn:AnsatzG}, that is fully compatible with the experimental data, that is, with $\Delta_l = 0$. In this section, we want to show that, within reasonable assumptions, the structure proposed in Eq.~\eqref{eqn:AnsatzG} can naturally arise in a broad class of models. Namely, those assumptions are: (i) the charged scalar field $\eta^+$, introduced in Eq.~\eqref{eqn:GeneralLagrangian}, can be associated with a neutral scalar field, and (ii) the corresponding scalar neutral interactions with the leptons both conserves flavor and generates the charged lepton masses.

To illustrate our arguments, we restrict ourselves to a model with the same fermion content as the SM but with $N$-Higgs doublets in the scalar sector. Generalizations to more complex models satisfying the two hypotheses above can easily be made. In this model, the most general Yukawa interactions for the leptons are given by
\begin{equation}
\label{eqn:NHiggs-Yukawa-Lepton}
    -\Lag_\text{Yukawa}^\text{Lepton} = \sum_{n=1}^N \lambda^n_{ab} \bar{\ell'}_{aR} \phi_n^\dagger L_{b} + \Hc,
\end{equation}
where $L_a = (\nu'_{aL}, \ell'_{aL})^T$ is the usual lepton doublet, and $\phi_n = (\varphi_n^+, \varphi_n^0)^T$ are the $N$ scalar doublets. The right-handed fields are the usual SM lepton singlets, and all fermionic fields are gauge symmetry states. Finally, the indices $a,b = 1,2,3$ label the different fermion generations.

From Eq.~\eqref{eqn:NHiggs-Yukawa-Lepton}, we see that if the neutral scalars develop a non-zero vacuum expectation value (VEV), $\ev{\varphi_n^0} = v_n/\sqrt{2}$, the (non-diagonal) charged lepton mass matrix is given by
\begin{equation}
\label{eqn:NHiggs-Lep-Mass-Matrix}
    M^{\ell} = \sum_{n} \frac{v_n}{\sqrt{2}} \lambda^n.
\end{equation}

As usual, this mass matrix can be diagonalized by two unitary transformations, as
\begin{equation}
\label{eqn:NHiggs-Lep-Mass-Matrix-Diag}
    \hat{M}^\ell = \left(V_R^\ell\right)^\dagger M^\ell V_L^\ell = \diag(m_e, m_\mu, m_\tau).
\end{equation}

Therefore, omitting the generation indices, the leptonic Yukawa interactions in Eq.~\eqref{eqn:NHiggs-Yukawa-Lepton} take the following form in the mass basis
\begin{equation}
\label{eqn:NHiggs-Lag-Yukawa-Mass}
    -\Lag_\text{Yukawa}^\text{Lepton} = \sum_n \left\{(\bar{\ell}_R X^n \nu_L)\varphi_n^- + (\bar{\ell}_R Y^n \ell_L)\varphi_n^{0*}\right\} + \Hc,
\end{equation}
where the mass states are given by $\nu_L = \left(V_L^\nu\right)^\dagger \nu'_L$ and $\ell_{L,R} = \left(V_{L,R}^\ell\right)^\dagger \ell'_{L,R}$, and we have defined
\begin{equation}
    X^n = \left(V_R^\ell\right)^\dagger \lambda^n V_L^\nu, \qand Y^n = \left(V_R^\ell\right)^\dagger \lambda^n V_L^\ell. 
\end{equation}

At this point, since the $\lambda^n$ coupling matrices, in general, are not individually diagonalized by the mass transformations of the charged leptons, flavor-changing neutral currents (FCNC) may occur in the lepton sector. However, from the current experimental results, we know that FCNC interactions must be heavily suppressed~\cite{Pich:2018njk,Oliveira:2022vjo}.

As shown by Glashow and Weinberg~\cite{Glashow:1976nt} and, independently, by Paschos~\cite{Paschos:1976ay}, a natural way to avoid these FCNC interactions in the scalar sector, at tree-level, is to assume that only one Higgs multiplet couples to each charged sector, due to an appropriately chosen discrete or continuous symmetry. This result is part of what is known in the literature as the Glashow-Weinberg-Paschos (GWP) theorem. In our case, we want to apply this result to the lepton sector. Then, assuming that this theorem holds, the only non-zero couplings in Eq.~\eqref{eqn:NHiggs-Lag-Yukawa-Mass} will be
\begin{equation}
\label{eqn:NHiggs-Yukawa-Mass}
    X = Y U, \qand Y = \sqrt{2} \frac{\hat{M}^{\ell}}{v_\ell},
\end{equation}
corresponding to the particular $\phi_n$ doublet that couples to the charged leptons, being $v_\ell$ its associated VEV, and $U = \left( V_L^\ell\right)^\dagger V_L^\nu$ the usual PMNS matrix. As the Yukawa coupling with the neutral scalar, $Y$, is now diagonal, FCNC does not appear at tree-level. The above results also hold in other situations where the GWP theorem does not. However, FCNC are still suppressed as, for example, in the so-called aligned models~\cite{Pich:2009sp,Botella:2015yfa,Enomoto:2015wbn,Gori:2017qwg,Penuelas:2017ikk,Alves:2017xmk,deMedeirosVarzielas:2019dyu,Diaz-Cruz:2020pjf}, where the Yukawa couplings are proportional one to another, or the Branco-Grimus-Lavoura (BGL) model~\cite{Branco:1996bq}, where the Yukawa entries are dependent only on CKM matrix elements and on the lepton masses~\cite{Branco:1996bq,Botella:2015yfa,Botella:2018gzy,Cornet-Gomez:2021lep}.

Apart from a possible scalar mixing, we can identify $\eta^-$ with the charged scalar present in the only doublet that couples to the charged leptons in Eq.~\eqref{eqn:NHiggs-Lag-Yukawa-Mass}. Then, substituting the results of Eq.~\eqref{eqn:NHiggs-Yukawa-Mass} in Eq.~\eqref{eqn:EffectiveScalarOperator}, we find that the effective coupling $G^\eta$ is given by
\begin{equation}
\label{eqn:G-Eta-GWP}
    G^\eta_{ij,lk} = \frac{\sqrt{2}}{m_\eta^2}\left(\frac{c^p_{ij}}{V_{ij}}\right) \left(\frac{m_l}{v_\ell}\right) U_{lk}, \qq{(no sum in $l$)} 
\end{equation}
where $c^p$ represents the pseudoscalar Yukawa couplings of the quark sector with the new scalar field, as in Eq.~\eqref{eqn:EffectiveScalarOperator}. Comparing with Eq.~\eqref{eqn:AnsatzG}, we have that
\begin{equation}
\label{eqn:GWP-Exact-Cancel-Lep}
    a_P = \sqrt{2}\frac{c^p}{m_\eta^2 v_\ell},
\end{equation}
in models that satisfy the GWP theorem in the lepton sector.

In this way, we have shown the existence of a broad class of models where the non-trivial solution given in Eq.~\eqref{eqn:AnsatzG} can naturally arise. Namely, in models that satisfy the GWP theorem and avoid FCNC in the lepton interactions mediated by scalars.

\subsection*{Applying the GWP theorem to the quark sector}

Although, as just shown, the main structure in Eq.~\eqref{eqn:AnsatzG} arises just by applying the GWP theorem to the lepton sector, even stronger bounds can be placed in FCNC from the quarks. For this reason, extending the above analysis to the quark sector is sensible. Considering, again, our simple N-Higgs model, the most general Yukawa interactions with quarks are
\begin{equation}
\label{eqn:NHiggs-Yukawa-Quark}
-\Lag_\text{Yukawa}^\text{Quark} = \sum_n \left\{\alpha^n_{ab} \bar{Q}_a \phi_n d'_{bR} + \beta^n_{ab} \bar{Q}_a \tilde{\phi}_n u'_{bR} \right\} + \Hc,
\end{equation}
where $Q_a = (u'_{aL}, d'_{aL})^T$ are the usual quark doublets, and $\tilde{\phi}_n = i\sigma_2 \phi_n^*$.

Again, by the GWP theorem, only one of the scalar doublets will give mass to each charged quark sector (but not necessarily the same for both). Then, the same steps that we followed for the leptons apply here, and the relevant Yukawa couplings of the quarks with the charged scalar are given by
\begin{equation}
\label{eqn:NHiggs-Yukawa-Quarks-GWP}
    X^d = \left(V_L^u\right)^\dagger \alpha V_R^d = \sqrt{2} \xi^d \frac{\hat{M}^d}{v_d} V, \qand X^u = \left(V_L^d\right)^\dagger \beta V_R^u = \sqrt{2} \xi^{u*} \frac{\hat{M}^u}{v_u} V^\dagger,
\end{equation}
where $V = (V^u_L)^\dagger V_R^d$ is the CKM matrix, $\hat{M}^u$ and $\hat{M}^d$ are the diagonal mass matrices for the up- and down-type quarks, respectively, $\alpha$ and $\beta$ are the (only) non-zero Yukawa that contributes to the quark masses, and $v_u, v_d$ are the VEVs associated with the scalars that couple to each corresponding charged sector. In the previous section, when we were applying the GWP theorem only on the lepton sector, we suppressed a possible scalar mixing factor, since it could always be absorbed in the free parameter $a_P$. Now, in principle, we can have one scalar for each charged sector, hence we must be more careful with those scalar mixing factors (that we represent by $\xi^u$, $\xi^d$, and $\xi^\ell$). Therefore, we take the Yukawa coupling with the lepton sector of Eq.~\eqref{eqn:NHiggs-Yukawa-Mass} as
\begin{equation}
    X = \sqrt{2} \xi^\ell \frac{\hat{M}^\ell}{v_\ell} U,
\end{equation}
that corresponds to the substitution $a_P \rightarrow \xi_\ell a_P$ in the Eq.~\eqref{eqn:G-Eta-GWP} above.

Comparing Eq.~\eqref{eqn:GeneralLagrangian} with Eq.~\eqref{eqn:NHiggs-Yukawa-Quark}, and using Eq.~\eqref{eqn:NHiggs-Yukawa-Quarks-GWP}, we have that the scalar and pseudoscalar coupling matrices, $c^s$ and $c^p$, respectively, are given by
\begin{equation}
    c^s = \xi^d X^d - \xi^u (X^u)^\dagger, \qand c^p = \xi^d X^d + \xi^u (X^u)^\dagger.
\end{equation} 

Finally, since only $c^p$ contribute at tree-level to the charge meson decays $P_{\ell 2}$, the effective coupling $G^\eta$ in this case is given by
\begin{equation}
\label{eqn:G-Eta-GWP-Full}
    G^\eta = \frac{1}{v^2 m_\eta^2} \left( \xi^d\hat{M}^d + \xi^u \hat{M}^u \right) \left(\xi^\ell \hat{M}^\ell U \right),
\end{equation}
where, for convenience, we have factored out the SM VEV $v \approx 246\,{\rm GeV}$, and rescaled the dimensionless scalar mixing factors $\xi^f$ by $\xi^f \rightarrow \xi^f v_f/v$ ($f=u,d,\ell$). Note that the effective coupling constant defined in Eq.~\eqref{eqn:G-Eta-GWP-Full} does not include the CKM matrix, as its elements were factored out on the effective Lagrangian in Eq.~\eqref{eqn:EffectiveLagrangian}.

Comparing Eq.~\eqref{eqn:G-Eta-GWP-Full} with Eq.~\eqref{eqn:AnsatzG}, we find that, for models where the theorem is valid in both the quark and lepton sectors,
\begin{equation}
\label{eqn:AP-GWP-Full}
    a_P  = b^u_P m_{u_i} + b^d_P m_{d_j},
\end{equation}
where $u_i$ and $d_j$ are the quarks present in the $P$ meson, and we have defined
\begin{equation}
    b_P^u=\frac{\xi^u\xi^\ell}{m_\eta^2 v^2}, \qquad b_P^d = \frac{\xi^d \xi^\ell}{m_\eta^2 v^2},
\end{equation}
for later convenience. Notice that the dimension of $b^{u,d}_P$ is different from that of $a_P$, due to the factorization of the quark masses.

\begin{figure*}[htb]
\centering
\includegraphics[width=\textwidth]{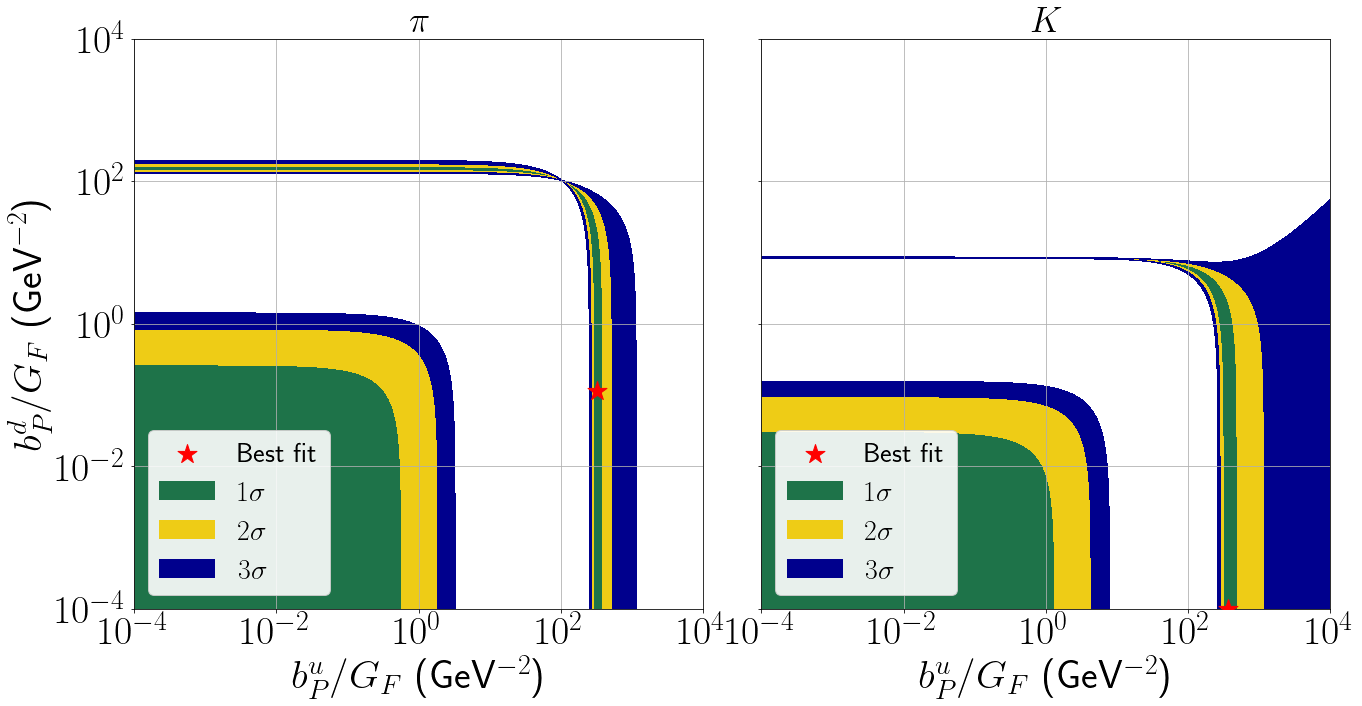}
\caption{Allowed region using pion (kaon) data left (right) considering the GWP mechanism in both the quark and the lepton sectors for the effective couplings $b_P^u/G_F$ and $b_P^d/G_F$.}
\label{fig:quarkGWP}
\end{figure*}
For $a_P$ real, we already saw that an exact non-trivial solution occurs for $a_P/G_F = 2/B_P$ (c.f. Eq.~\eqref{eqn:Real-AP-Solution}). Now, using Eqs.~\eqref{eqn:FactorBP} and~\eqref{eqn:AP-GWP-Full}, we can rewrite this solution as
\begin{equation}
\label{eqn:AP-GWP-Full-2}
    \frac{1}{G_F}(b^u_P m_{u_i} + b^d_P m_{d_j}) = \frac{2}{m_P^2}(m_{u_i} + m_{d_j}).
\end{equation}
Moreover, we see that a similar structure in the quark masses appears in both sides of Eq.~\eqref{eqn:AP-GWP-Full-2}.

\begin{figure*}[htb]
\centering
\includegraphics[width=0.9\textwidth]{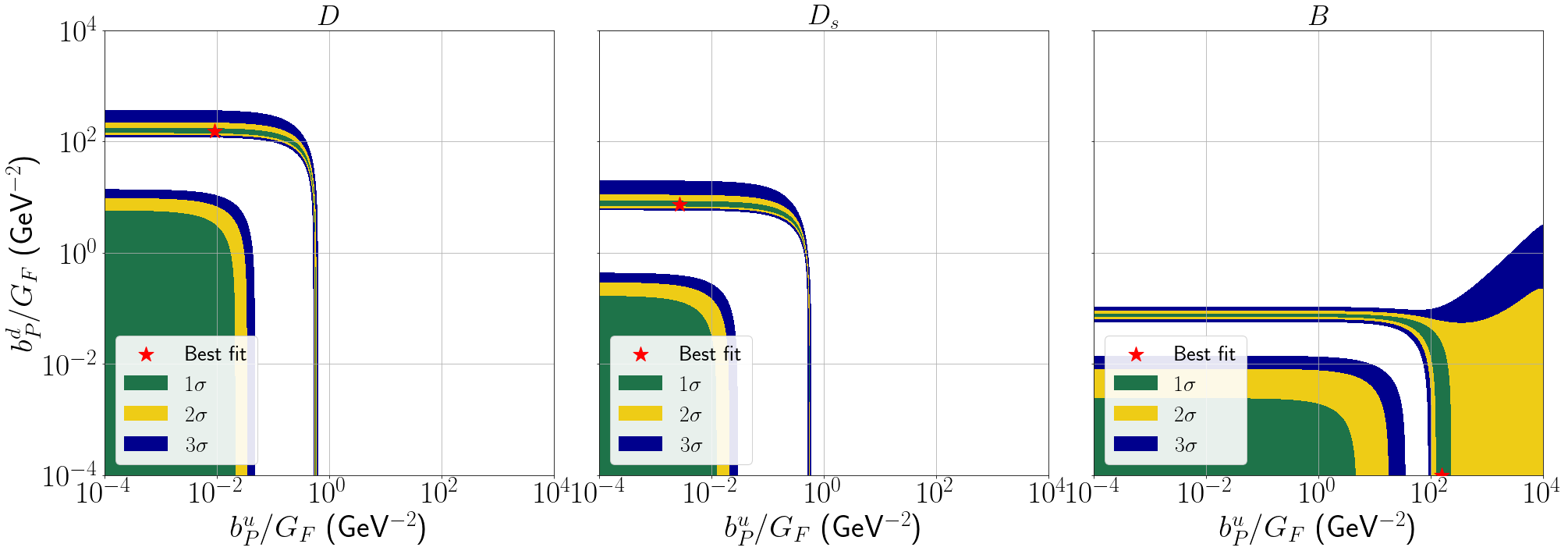}
\caption{The same as Figure~\ref{fig:quarkGWP} for the heavy mesons $D$ (left), $D_s$ (center) and $B$ (right).}
\label{fig:quarkGWPHeavy}
\end{figure*}

The allowed parameter space for this case is shown in Figure~\ref{fig:quarkGWP}, for the most precise measurements of $\pi_{\ell 2}$ and $K_{\ell 2}$. Figure~\ref{fig:quarkGWPHeavy} shows the same analysis for the heavier mesons $D$, $D_s$, and $B$. Now, three different situations can occur (i) both $b^u_P \approx b^d_P \approx 0$, (ii) only one of the parameters $b^{u,d}_P$ are non-zero, and (iii) both parameters are non-zero. The first situation corresponds just to an SM-compatible solution, where the new physics is small compared to the SM contribution. In the second case, we are in a similar situation as that given in Figure~\ref{fig:Real-AP-Mesons}, and the non-trivial solution is given in Eq.~\eqref{eqn:Real-AP-Solution} with the replacement $a_P \rightarrow b^u_P m_{u_i}, b^d_P m_{d_j}$. Comparing the allowed regions for pions and kaons in Figure~\ref{fig:quarkGWP}, we see that, for $b^u_P \neq 0$ and $b^d_P \rightarrow 0$, both corresponds to a similar range, and we have that $b^u_\pi = b^u_K$ is a possible solution; while for the opposite case, $b^u_P \rightarrow 0$ and $b^d_P \neq 0$, the asymptotic value for $b^d_P$ is very different, and no solution of the form $b^d_\pi = b^d_K$ can occur. This happens due to the quark content of those mesons, while they share the same up-type quark content, the down-type quark is different, with a down quark for the pions and a strange quark in the kaons. Finally, for case (iii) above, we have the full solution given in Eq.~\eqref{eqn:AP-GWP-Full-2}, and it differ for each meson $P$.

Again, no universal solution in the quark sector is possible for the parameters $b_P^u$ and $b_P^d$ in the non-trivial region. However, in this non-trivial region, as it is possible to see from Eq.~\eqref{eqn:AP-GWP-Full} and the almost constant factor $B_P$ for all mesons, couplings related to the same quark asymptotic to the same values, e.g., $\pi$ and $K$ coupling $b_P^u$, which is related to the up-quark go to the same value when $b_P^d \rightarrow 0$. This result suggests that non-universality is preferred for quarks in this region, but couplings related to the same generation of quarks are consistent among different mesons. Remembering that $B_P$ is not so close to the value of the ratio for the other mesons, we can also understand why $b_P^u$ asymptotic to a slightly different value for $B$ than for $\pi$ and $K$. We also point out that the parameters $a_P$ and $b^{u,d}_P$ have different dimensions since the quark masses have been factored out in Eq.~\eqref{eqn:AP-GWP-Full}.

\section{Constraints on the pseudoscalar mass}
\label{sec:Mass-Constraints}

This section uses experimental and theoretical constraints to restrict the charged Higgs mass in this new region. For the lower limit, we use current searches for charged scalars on collider experiments, specifically from the Large Electron-Positron (LEP) collider experiment~\cite{Workman:2022ynf}. We use the perturbative limits on the effective Lagrangian couplings for the upper limit on the Higgs mass.

Assuming that the quark couplings in Eq.~\eqref{eqn:GWP-Exact-Cancel-Lep}, $c^p$, respects the perturbative limits, that is $c^p < \sqrt{4\pi}$, we have that
\begin{equation}
    \frac{a^P}{G_F} \equiv \frac{c^p}{m_\eta^2 v_\ell G_F} < \frac{\sqrt{4\pi}}{m_\eta^2 v_\ell G_F}.
\end{equation}
Hence, we must have
\begin{equation}
    m_\eta < \sqrt{\frac{\sqrt{4\pi}}{(a^P/G_F)v_\ell G_F}}
\end{equation}

Now, for example, if we use the best-fit point for $\pi_{\ell 2}$ in Figure~\ref{fig:Real-AP-Mesons}, $a^P/G_F = 0.7\,{\rm GeV}^{-1}$, we have that
\begin{equation}
    m_\eta < \left(658.33 \, \sqrt{\left(\frac{1 \, {\rm GeV}}{v_\ell}\right)}\right) {\rm GeV}
\end{equation}

If we take $v_\ell = v_{\rm SM}$, we have that
\begin{equation}
    m_\eta < 42\,{\rm GeV},
\end{equation}
and this case is already excluded by the LEP low limit of $80\,{\rm GeV}$.

But, if we use $v_\ell = 2\,{\rm GeV}$, corresponding to the  perturbative limit necessary for the tau mass, we have that 
\begin{equation}
    m_\eta < 466\,{\rm GeV}.
\end{equation}

Therefore, we found that the scalar mass should be in the range
\begin{equation}
    80\,(181)\,{\rm GeV} < m_\eta < 466\,{\rm GeV},
\end{equation}
where the lower bound comes from charged scalar searches at LEP as reported on PDG~\cite{Workman:2022ynf}.

\section{Conclusions}

In this work, we studied the leptonic decays of charged pseudoscalar mesons, $P_{\ell 2}$, which is mediated by a novel scalar field. Although such decays have already been widely analyzed in the literature, we have proposed a new non-trivial solution, given in Eq.~\eqref{eqn:AnsatzG}, that depends only on one free parameter, $a_P$, for each meson. We have shown that such a solution can altogether avoid the constraints coming from the ratio $R_{l/l'} = \Gamma(P \rightarrow l\nu[\gamma])/\Gamma(P \rightarrow l'\nu[\gamma])$ and still allows a relatively large region in parameter space when the individual decay rates are used in the statistical analysis. As can be seen in Figs.~\ref{fig:Real-AP-Mesons} and~\ref{fig:Complex-AP-Pion}, a larger allowed region appears beyond the fine-tuned non-trivial solution. Taking $\pi_{e2}$ as an example, we see that our proposed solution permits an effective coupling for the new contribution in the range of $10^{-4}\lesssim (G^\eta/G_F) \lesssim 10^{-3}$, as can be seen in Figure~\ref{fig:Real-AP-Mesons}. On the other hand, we also have shown that no universal solution for all the pseudoscalar mesons, that is, $a_\pi = a_K = a_D = a_{D_s} = a_B$, could be found.

Moreover, we have shown that such a non-trivial solution can naturally emerge in models where flavor-changing neutral currents in the scalar sector are avoided in interactions with leptons. The most notorious examples are the models that satisfy the so-called Glashow-Weinberg-Paschos theorem. However, our results remain valid in other models where FCNC are avoided, such as, for example, the aligned models or the Branco-Grimus-Lavoura model. We then study the case where the FCNC in the scalar sector is avoided in interactions with quarks and leptons. Finally, we have also estimated the mass of the new scalar to be in the range $80\, (181) \text{ GeV} < m_\eta < 466$~GeV.

To conclude, the non-trivial solution may also impact other relevant physical processes, for example, beta decay, the semi-leptonic meson decays, and tau and muon decays. Although such processes are out of the scope of this current work, to consolidate this solution as a candidate for new physics or to exclude it definitively, their analysis will be done in the future.

\section*{Acknowledgments}
 We are thankful for the support of FAPESP funding Grant 2014/19164-6. In addition, OLGP is thankful for the support of CNPq grant 306565/2019-6 and 306405/2022-9, and SWPN is also thankful for the support of CNPq grant 140727/2019-1. LJFL is grateful for CAPES and CNPq support, grants 88887.613742/2021-00 and 131548/2019-0. Furthermore, this study was financed in part by the Coordenação de Aperfeiçoamento de Pessoal de Nível Superior - Brasil (CAPES) - Finance Code 001. Finally, we thank a previous referee that challenged us to understand the two solutions found in this paper.

\appendix
\section{Statistical analysis}
\label{sec:StatisticalAnalysis}

As mentioned, the ratio $R_{l/l'}$ given in Eq.~\eqref{eqn:Br-Ratio} is unsuitable for statistical analysis for the particular structure of Eq.~\eqref{eqn:AnsatzG}, as it will always cancel out the new physics terms. With that in mind, we use the individual decay rates for each decay channel to fit the experimental data to our calculated rates.

The statistical analyses were performed using the following definition of the $\chi^2$ function,
\begin{eqnarray}{\label{eqn:Chi-Squared-Test-Rate}}
    \chi^2(x) = \sum_{l}\frac{\left(\Gamma_l(x) - \Gamma_l^{\rm exp}\right)^2}{\left(\sigma^{\rm SM}_{\Gamma_l}\right)^2+\left(\sigma_{\Gamma_l}^{\rm exp}\right)^2+\left(\sigma^{\eta}_{\Gamma_l}(x)\right)^2},
\end{eqnarray}
where $\sigma^{\rm SM}_{\Gamma_l}$ is the uncertainty in SM theoretical calculations for the leptonic meson rate given in Eq.~(\ref{eqn:Total-Decay-Rate}), $\sigma^{\eta}_{\Gamma_l}$ is the propagated uncertainty in the new physics terms due the charged scalar, and $\sigma_{\Gamma_l}^{\rm exp}$ is the experimental uncertainty of decay rate,  $\Gamma_l^{\rm exp}$~\cite{Workman:2022ynf}. Finally, we have $x=a_P/G_F$ for the analysis present in Figure~\ref{fig:Real-AP-Mesons}, $x=a^r_P/G_F,\,a^i_P/G_F$ for Figure~\ref{fig:Complex-AP-Pion} and $x=b^d_P/G_F,\,b^u_P/G_F$ for Figures~\ref{fig:quarkGWP} and \ref{fig:quarkGWPHeavy}. We use the confidence regions for one free parameter for the real case and two free parameters for the complex and the quark sector GWP structure.

\end{document}